# The Interplay Between Imprint, Wake-Up Like Effects and Domains in Ferroelectric $Al_{0.70}Sc_{0.30}N$

*Maike Gremmel\* and Simon Fichtner*

Simon Fichtner, Maike Gremmel.
Department of Material Science, Kiel University, Kaiserstraße 2, 24143 Kiel, Germany
Email: magr@tf.uni-kiel.de

Simon Fichtner
Fraunhofer Institute of Silicon Technology (ISIT), Fraunhoferstraße 1, 25524 Itzehoe, Germany
Email: simon.fichtner@isit.fraunhofer.de

**Abstract:**

This paper investigates wake-up and imprint in ferroelectric $Al_{0.70}Sc_{0.30}N$ films. The study employs a series of I-V and P-E measurements with varying electric field amplitudes and voltage cycles as well as structural investigation via Scanning Electron Microscopy to understand the origin and underlying principle of wake-up and imprint as well as their relation. It is shown that the material can be considered wake-up free, however inherent imprint and imprint shift in combination with minor loops result in a wake-up like effect. We introduce a proposition to explain the influence of initial switching cycles on domains, their stabilization and corresponding changes in imprint. Unipolar fields and temperature investigations are used to explore the reversibility of imprint and ways to program it, while partial switching is applied to investigate domain propagation and support the aforementioned approach. It is concluded, that after an energetically more demanding domain nucleation, domain wall motion can switch the majority of polarization in $Al_{1-x}Sc_xN$. As a consequence, the presence of initial domains reduces the coercive field in respect to unipolar films.

# 1. Introduction

Ferroelectricity is defined as the ability to switch the spontaneous electric polarization present in certain materials with a non-centrosymmetric crystal structure with the help of an external electric field. This property has been found to offer promising solutions in various fields such as non-volatile information storage, sensors and actuators.[1–3] As a field driven effect, ferroelectricity allows particularly energy efficient device operation.[3]

$Al_{1-x}Sc_xN$ and related wurtzite-type solid solutions are a novel class of ferroelectric thin film materials that offer huge spontaneous polarizations, very good long term stability, as well as complementary metal-oxide-semiconductor (CMOS) compatibility.[4–6] This makes them attractive candidates for use in microelectronic devices, where such properties are often critical. The extraordinary temperature stability of $Al_{1-x}Sc_xN$ also allows for its use in high-temperature applications, [4] while the high coercive field ($E_c$) of $Al_{1-x}Sc_xN$ also enables its scalability to very thin layers while maintaining sufficient resistance to depolarization.[7–9]

These properties of wurtzite-type ferroelectrics make them a potent solution for use in a variety of applications, including compact and low-power electronics, sensors, and actuators.

In the context of bringing wurtzite-type ferroelectrics like $Al_{1-x}Sc_xN$ to applications, a comprehensive understanding of common phenomena in ferroelectrics will be required. This includes the wake-up and imprint of their polarization vs. electric field (P-E) hysteresis loops. While imprint is defined as an asymmetric ferroelectric hysteresis loop where one polarization state is preferred over the other, [10–13] wake-up describes the increase of remanent Polarization $P_r$ over a number of switching cycles.[14] Both effects are in principle undesired properties that can lead to a device failure and therefor have to be characterized and addressed. In this paper, we highlight the hitherto unreported connection between wake-up like effects, imprint and domain propagation in wake-up free $Al_{1-x}Sc_xN$ films.

While studies on wake-up in fluorite- and perovskite-type ferroelectric films have already provided a good understanding of processes inside these materials, little is known to date about wake-up in the novel wurtzite type ferroelectrics.[15–18] Wake-up was reported on recently by different groups for this new material class, albeit with no general behavior or mechanism available to date.[19–23] One possible mechanism is introduced by Zhu et al. for $Al_{1-x}B_xN$, where they attribute wake-up to the nucleation and growth of stable inversion domains.[19] By repeatedly switching the films, the number of stable inversion domains increases, which lead to a higher volume fraction getting switched in subsequent cycles. Therefor the remanent polarization $P_r$ increases up to the point where the full volume is inverted. On the other hand $Al_{1-x}Sc_xN$ was previously claimed to be wake-up free, albeit without in-depth analysis.[24] Therefore, no clear understanding of the general mechanisms of wake-up like behavior in wurtzite-type thin films exists to date, while the same applies for the imprint which is often found in P-E loops of these films.[25]

Therefore the aim of this work will be to answer the question if, first of all, wake-up in our sputter-deposited $Al_{1-x}Sc_xN$ films exists, how the imprint evolves, what could be at their origin and how do they compare with previous reports from the class of the wurtzite-type ferroelectrics as well as ferroelectrics in general.

This work is separated into three sections. In the first, wake-up is investigated on $Al_{70}Sc_{30}N$ films via current density vs. electric field (I-V) measurements and the results are compared to $Al_{1-x}B_xN$. Further, the evolution of imprint up to $10^4$ switching cycles is analysed. In the second part, the influence of the inversion domains on the electrical response/ I-V peaks is investigated, supported by scanning electron microscopy (SEM) polarity investigations after wet etching. An explanatory model for the interplay between imprint, wake-up like effects and domains will be specified and delimited from the approach for $Al_{1-x}B_xN$. In a third part, thermal annealing is applied to investigate possible domain growth and the influence on imprint.

## 2. Experimental

For the investigation of wake-up, capacitor structures with 420 nm thick $Al_{1-x}Sc_xN$ films where prepared on oxidized Si (100) covered by a 100 nm Mo bottom electrode on a 100 nm AlN seedlayer.

The $Al_{1-x}Sc_xN$ films were grown in an Evatec MSQ 200 magnetron sputter tool by cosputtering from two single element targets. The Sc concentration was measured to be 30 at% by SEM EDX. Details regarding the deposition process were published elsewhere.[26] As top electrode, 100 nm Mo was used and structured by a mixture of phosphoric and sulfuric acid (PWS).[27]

For electrical characterization, current over electric field loops were measured with a triangular voltage signal without prepolarization in an aixACCT TF 2000. Polarization over electric field loops were obtained through integration of these loops. Coercive fields were determined by a fit of the I-V peaks (See supporting information Figure S4). PUND measurements (Positive up negative down) for leakage compensation were omitted, as the additional voltage pulses cause a distortion of the coercive fields, i.e. have a direct effect on the observed evaluation of the hysteresis loops.

For additional structural investigations the capacitor structures were etched in hot (80°C) phosphoric acid (80% $H_3PO_4$: 20% $H_2O$) at different states of cycling and investigated with scanning electron microscopy (SEM, Zeiss Ultra 55 Plus). Futhermore, X-ray diffraction on similar samples was performed in a previous study.[4]
Thermal annealing in a Nabertherm muffle furnace LTS/11 was used for investigation of domain growth and imprint shift at elevated temperature.

## 3. Results and Discussion

### 3.1 Wake-Up Like Effects and Imprint in $Al_{1-x}Sc_xN$

Initially, we discuss under which measurement conditions our ferroelectric $Al_{0.70}Sc_{0.30}N$ films feature wake-up like effects, while in general they can be considered wake-up free. Here, we follow the standard definition of wake-up as an increasing remanent Polarization ($P_r$) over the number of voltage cycles the material has seen. [10–13]

First, a series of I-V measurements with 30 cycles was performed with an electric field amplitude of 4.7 MVcm$^{-1}$. For an amplitude of 4.7 MVcm$^{-1}$, $P_r$ in the first switching cycle is virtually equal to the $P_r$ after 30 cycles, therefor the film can be considered wake-up free (**Figure 1a,** positive fields). However a negative imprint ($E_c^+ < |E_c^-|$) of all hysteresis loops is visible, where $E_c+$ is the positive coercive field and $E_c-$ the negative respectively. When the sum of positive and negative coercive field is plotted against the number of cycles, a clear decrease of this sum becomes obvious (Figure 1c). $E_c^+$ is slightly increasing during the first 20 cycles before it starts to decrease exponentially around 100 cycles (**Figure 2**). $|E_c^-|$ decreases exponentially right from the start up to $10^4$ cycles and with a larger exponent than $E_c^+$. This leads to a narrower hysteresis and constant reduction of imprint up to the point, where both coercive fields are nearly the same.

Measurements at only < 5% lower voltage amplitudes instead show a strong increase of $P_r$ in the initial cycles which can appear as a classical wake-up effect without measurements at higher fields. The lower the amplitude, the more cycles are necessary to wake-up the material, similarly to that was reported by Zhu *et al.*[19] From the P-E loops in Figure 1d it becomes evident, that the hysteresis loops of the cycles during wake-up are not saturated for negative fields while the positive side clearly is. This, and the $E_c^-$ shift implies that these measurements correspond to minor loops relative to $E_c^-$. With an additional bias field of -0.4 MVcm$^{-1}$, and thus a voltage signal which is more overlapping with the initial form of the P-E curve, it is again possible to reach full polarization without wake-up. The shift of the negative coercive field as present in the measurements at higher field appears as well. A comparison of $P_r$ for samples experiencing wake-up and biased ones is displayed in Figure 1e. All corresponding I-V loops are included in the supporting information **Figure S1**. While the biased measurement reaches 120 µCcm$^{-2}$ in the first cycle, without it the measurement only approaches this value after 10 cycles (Figure 1f). Due to the proximity to the breakdown field of the films, the data contains an unneglectable amount of leakage, leading to an overestimation of the true remanent polarization of up to 30μCcm$^{-2}$.

We interpret these results such, that initial minor loops gradually transform into saturated loops mainly through an increasing $E_c^-$. This increase in $E_c^-$ then shifts the hysteresis loop inside the probed voltage interval. Since virtually the whole volume of the film can however be switched in the first loop, provided a high enough voltage is selected, and a saturated $P_r$ is available in this loop, the term wake-up free can be applied to our Al$_{1-x}$Sc$_x$N films.

This nonetheless raises the question what is the origin of the imprint and the varying coercive fields we observe, which are responsible for the wake-up like behavior of the minor loops.

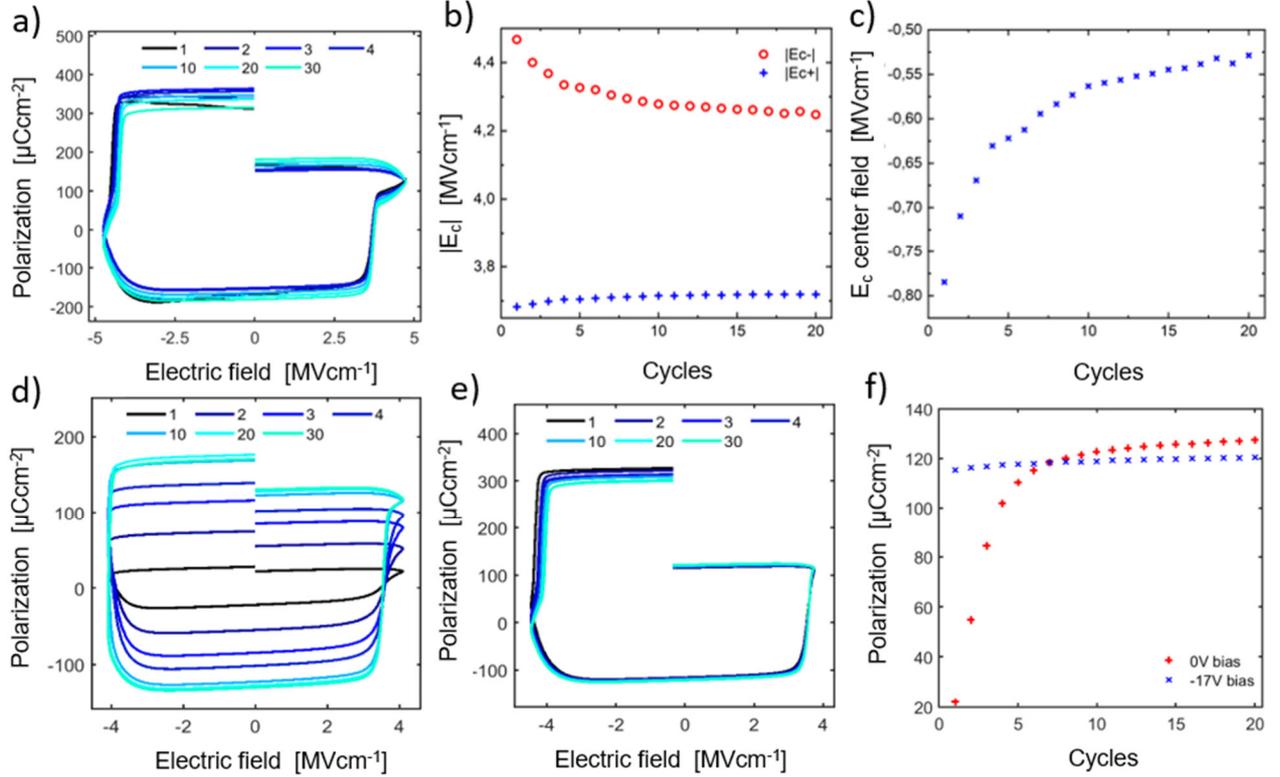

**Figure 1** : *a) P-E loops at 4.7 MVcm$^{-1}$ and a frequency of 211 Hz. b) Corresponding shift of $E_c^+$ and $E_c^-$ over the first 20 cycles on a pristine sample. In c) the sum ($E_c^+ + E_c^-$) is shown as a representation of the initial imprint shift calculated from values of b. d) P-E loops at 4.1 MV/cm and 211Hz without and e) with a bias field of -0.36 MVcm$^{-1}$ for the first 30 cycles on a pristine capacitor on the same sample. F) $P_r^+$ over cycles for sample exhibiting wake-up like behavior and the biased sample.*

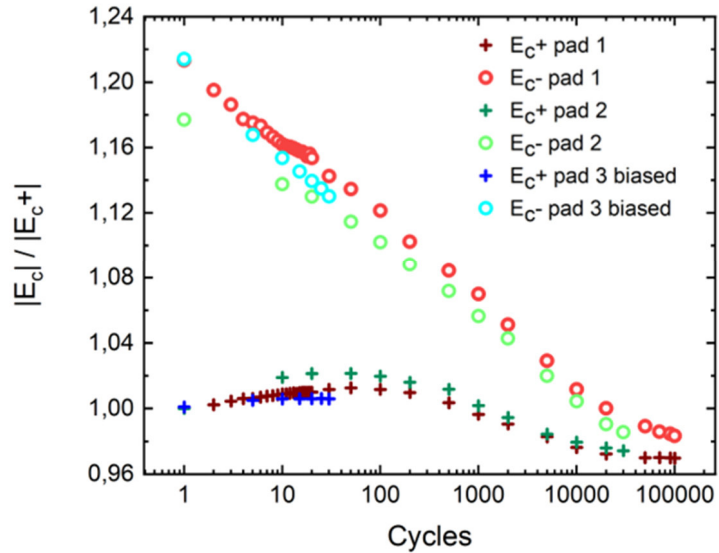

**Figure 2 :** *Positive and negative coercive field at 211 Hz normalized by the initial $E_c^+$ for different samples. Pad 1 and 3 correspond to Figure 1 a-c) and e) accordingly.*

We consider a similar explanation as given by Zhu *et al.* for $Al_{1-x}B_xN$ (formation of permanent inversion domains), to be responsible for the variation of $E_c$ and therefore also the wake-up like effects as well as the imprint[19]. Unlike for $Al_{1-x}B_xN$, we do however not consider these regions to necessarily correspond to real inversion domains in all cases (especially for the initial minority polarization), but as more general defect sites that allow nucleation of the initial minority polarization at reduced energies. In general wurtzite $Al_{1-x}Sc_xN$ can exhibit two polarities (**Figure 3**). In respect to the Al and Sc atoms the three equivalent Nitrogen bonds can either point down, towards the substrate, described as Metal-polar (M-polar) or up, towards the growth direction, described as Nitrogen polar ( N-polar). Consequently, during repeated switching, the originally exclusively N-polar film, starts to acquire more and more regions were the nucleation of M-polarity is energetically more favorable. These regions then serve as nucleation sites for the subsequent switching events from N- to M-polarity. Thus, they reduce the absolute value of $E_c^-$, by allowing energetically less expensive domain wall motion. This mechanism is sketched in **Figure 4**. The smaller shift of $E_c^+$ in comparison to $E_c^-$ is attributed to the as grown N-polarity which therefor is already energetically favored from the beginning and switching does not improve as much.

While Zhu *et al.* consider stable inversion domains to grow from top and bottom electrode for $E > E_c$ and remain there even after switching back, we did not find direct evidence for the existence of stable M polar residues after switching to N polarity at the film surface.[14] This was investigated by wet etching. Capacitors with different switching states were submerged in Chromium etchant for about 1 minute to remove the Mo electrode. Subsequently, the $Al_{1-x}Sc_xN$ layer was etched in $H_3PO_4$ at 80°C for 1 minute. While N-polar $Al_{1-x}Sc_xN$ etches in $H_3PO_4$, M-polar films are virtually inert.[25] Despite a line along the former electrode edge, the surface for fully N-polar capacitors in as deposited state and capacitors cycles >70 times appear the same. No remaining M-polar inversion domains are visible on the switched regions (**Error! Reference source not found.** 3). For M-polar capacitors, a clear interface towards the surrounding N-polar region is observed, again demonstrating the general feasibility to determine the local surface polarity of wurtzite-type films. Since we did not observe remaining M-polar regions on the top of films switched to positive saturation, we conclude that in N-polar state no M-polar domains remain at the electrodes. Instead, switching likely introduces regions where the nucleation of M-polarity becomes energetically more favorable resulting in the lowered coercive field and therefore reduced imprint. These regions are most likely situated under the upper electrode, as etching a film switched to only about 50% M-polarity yielded a virtually closed M-polar surface (Figure 3), strongly suggesting that switching originates from the upper interface. After apparently saturating M-polarity, we did however see indications for real residual N-polar domains at the bottom of the films, which can serve as starting points for domain propagation while back-switching. In our previous work, we could image such regions via TEM polarity determination and also conclude on their presence indirectly through a reduced piezoelectric coefficient.[28] Their presence might also explain why the N-polar polarity remains favorable during 1000's of cycles (Figure 2).

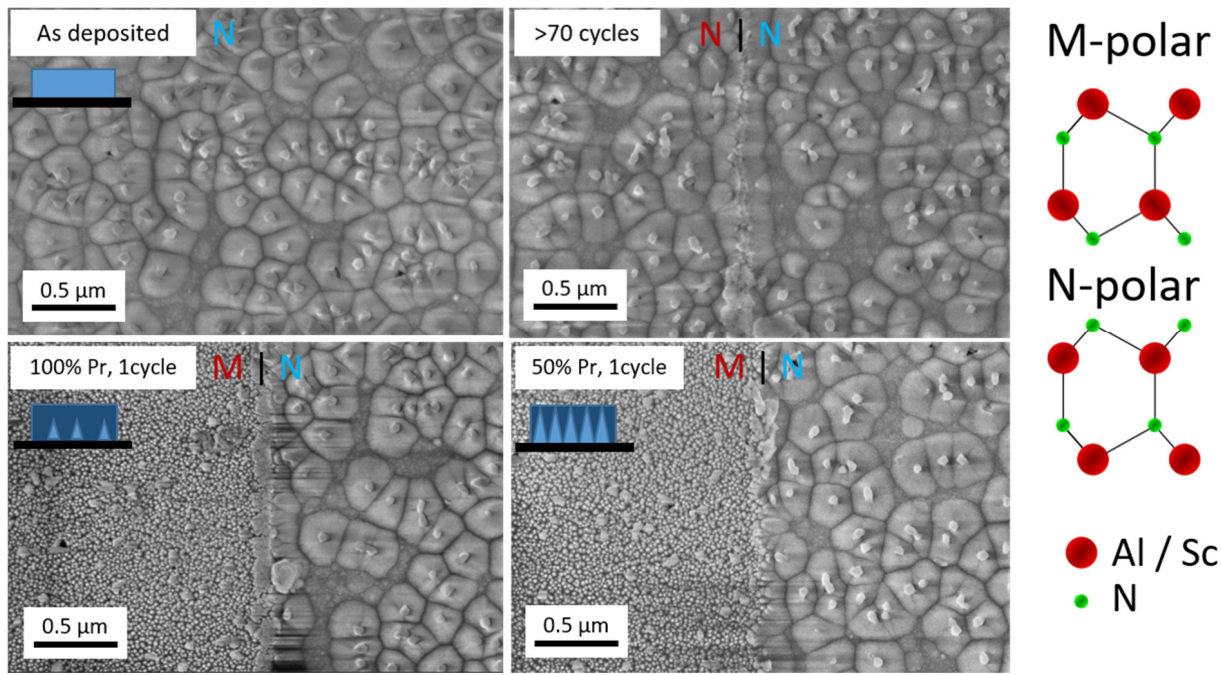

**Figure 3:** *SEM images of N polar and M polar capacitors after removing the electrode and anisotropic etching of the surface. Top left: as deposited capacitor, N-polar, Top right: > 70 cycles N-polar, bottom left: 1 cycle 100% $P_r$, M-polar, bottom right: 1 cycle < 50% $P_r$. M-polar.*

We therefore propose the following explanation for the initial switching cycles after deposition: During the first transition from N- to M-polarity, switching begins with nucleation at the top electrode, which then progresses through the film towards the bottom electrode by domain wall motion, where typically N-polar residuals remain. These serve as initial sites from which domain wall propagation switches the film back from M- to N-polarity, while regions more favorable for the nucleation of M-polar unit cells remain at the top. That the initial presence of small minority domains can have a profound impact on $E_c$ and thus the energy required to initiate switching, is demonstrated in detail in the following section.

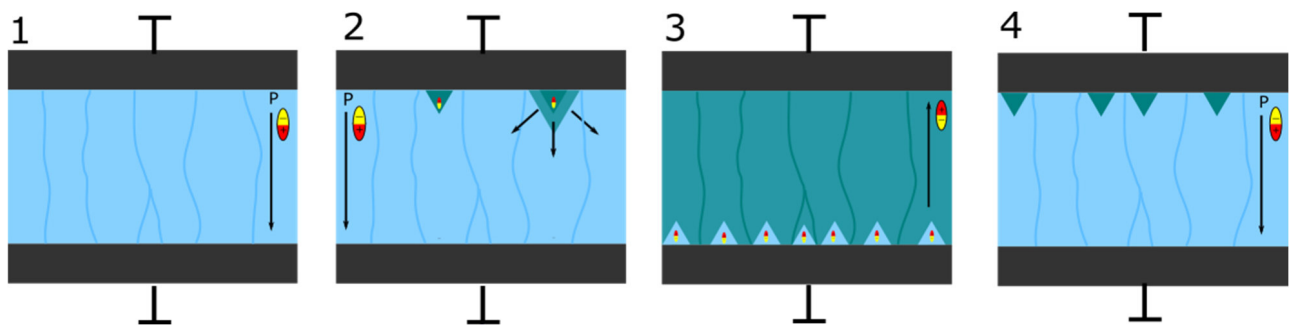

**Figure 4:** *Sketch for domain nucleation and propagation in ferroelectric $Al_{1-x}Sc_xN$.*
*1 - capacitor in as deposited N-polar state, 2 - nucleation and growth of M-polar domains under the influence of E, 3 - fully switched film at $P_r^-$, 4 - fully switched film at $P_r^+$ at >1 cycle.*

To investigate whether the stabilized imprint after the first cycles is reversible, I-V curves were measured at 211 Hz before and after 5 unipolar cycles (0.1 Hz) with all voltage peaks of either exclusively positive or negative polarity. Applications of a negative field to a sample that was cycled 40 times resulted in a shift of the peaks and a reduced imprint as shown in **Figure 5**. This shift was reversible by an introduction of a positive field. Similar effects are typically explained by charge injection, migration of mobile charges or dipolar defect rotation.[11,14] For $Al_{1-x}Sc_xN$ these mechanisms are however not able to fully explain the variation of the IV response. In first approximation a charge distribution should affect both coercive fields equally, yet the shift of the positive field is stronger pronounced. For charge migration one can further argue that starting from a negative imprint, either a higher positive carrier density is present at the top electrode, or a negative carrier density at the bottom electrode. A positive field in both cases would therefor lead to a more homogeneous distribution of the defects, reducing the imprint. However the imprint shifts in the opposite direction, bringing us to the conclusion, that migration of charges is not the driving force. A sketch of the mechanism is added in the supporting information **Figure S3**.

A similar argumentation is true for the dipolar defects: The internal field would only allow for the positive end of the majority of the dipole defects to face towards the top electrode. Therefore a positive applied field should rotate these, reducing the internal field and a negative applied field should increase the imprint. This is again opposite to the observations of this work and can therefore be ruled out as the driving mechanism.

In contrast to that, the direction of the shifts is coherent with our proposed explanation of domain growth which also provides a reasonable explanation as to why both coercive fields do not shift in unison. When a negative field is applied to the sample, the N-polar film is switched to M-polarity. Keeping the field applied reduces the residual N-polar inversion domains and in return N polarity requires higher fields to switch in subsequent cycles. Therefore the hysteresis is shifted to lower imprint, especially on the positive side. For a positive field, the stabilization is reversed. Switching to M-polarity requires larger fields and the hysteresis shifts back to a higher imprint.

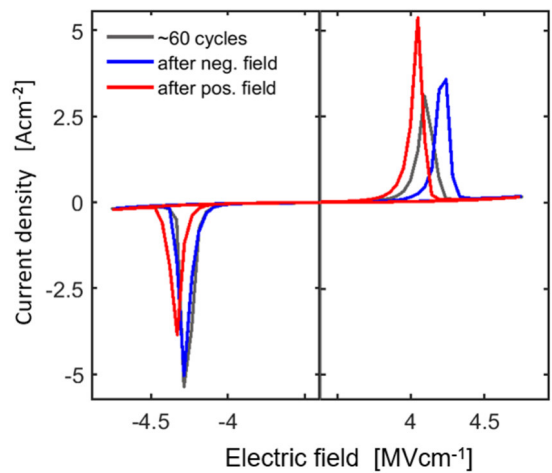

**Figure 5:** *I-V measurements of the $Al_{0.70}Sc_{0.30}N$ film at 211 Hz. A triangular unipolar field of -4.7 $MVcm^{-1}$ was applied 5 times at 0.1 Hz before measuring the blue curve. Afterwards the same was done with a positive amplitude corresponding to the red plot.*

## 3.2 Nucleation Trough Partial Switching and its Effect on $E_C$

To further support our interpretation of the origins of imprint and the resulting wake-up like behavior of minor P-E loops in $Al_{1-x}Sc_xN$, we investigated the interplay between nucleation and the continuation of polarization switching in more detail. The influence of nucleation and the subsequent growth of inversion domains on the electrical properties becomes especially clear in the observation of I-V peaks during and after partial unipolar switching. In the first cycle a minor loop (90% of $E_c$) was applied to partially switch an N-polar film to M-polarity (or vice-versa, the switching polarization was kept between 5% and 25% of $2P_r$ for each case). During the second pulse, the previously partially switched polarity was saturated. A third pulse was used to fully switch the sample back to its initial polarity and the forth pulse to reach saturation in M-polarity (or N-polarity) again in a single step to determine the full polarization. These measurements were conducted on both pristine and repeatedly switched samples as well as under different frequencies between 50 Hz and 700 Hz.

As shown in **Figure 6,** the current peaks of the minor loop (1$^{st}$ cycle) as well as the peak from the 4$^{th}$ cycle (after full inversion to M polarity) start to increase at a field of about 3.5 MVcm$^{-1}$ with the same slope. The 2$^{nd}$ cycle (full cycle after minor loop) however starts at 3 MVcm$^{-1}$, right where the 1$^{st}$ peak recedes to. Additionally, the 2$^{nd}$ peak is broader, consequently has a lower current density and the coercive field is reduced as well. Measurements at different frequencies and for both polarities give comparable results as shown in the supplement **Figure S2**.

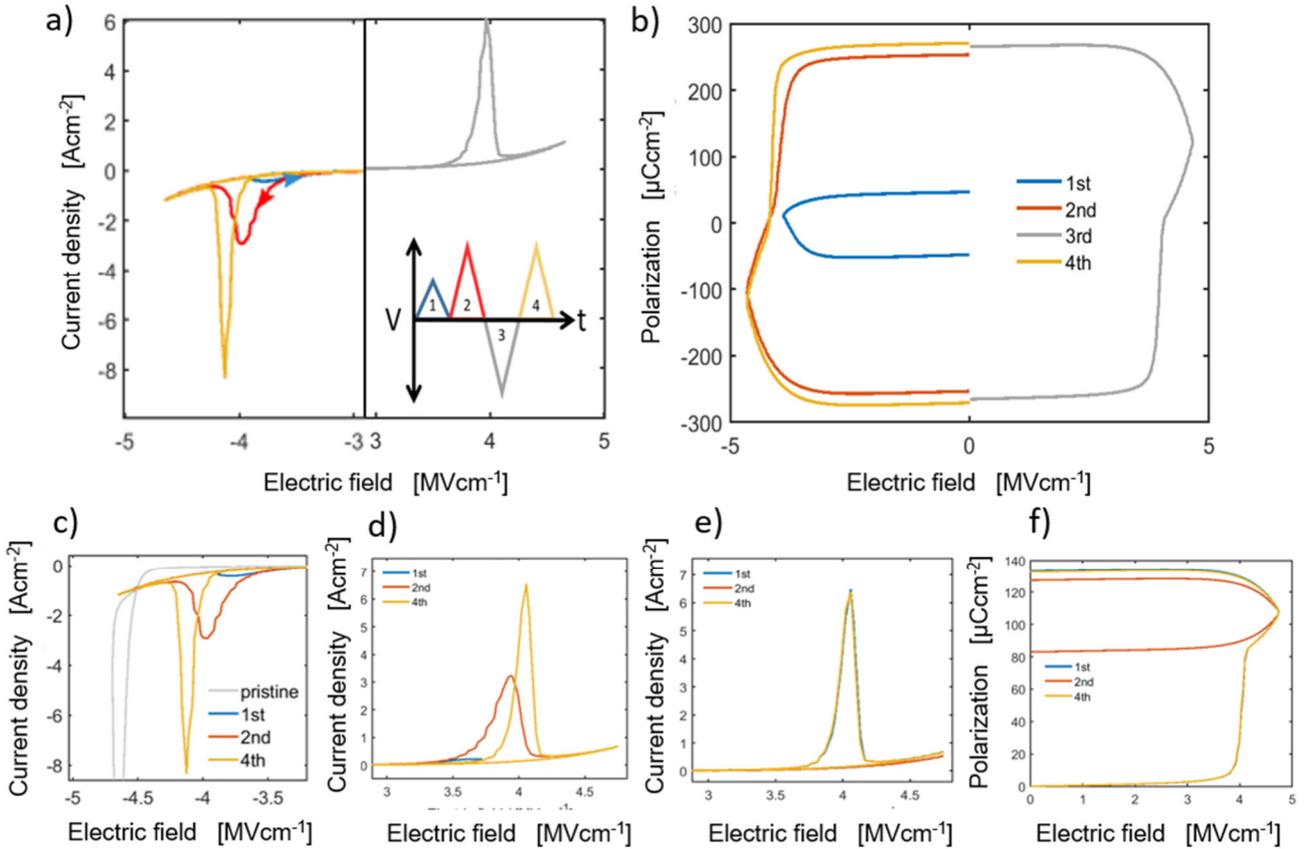

**Figure 6:** *a) I-V measurement with previous nucleation cycle, pulse train and corresponding b) P-E measurements at 211 Hz. Blue and red are subsequent unipolar measurements with triangular voltage amplitude of -3.9 MVcm$^{-1}$ and -4.6 MVcm$^{-1}$ respectively. The yellow and gray curves were also measured with*

-4.6 MVcm$^{-1}$. c, d) I-V measurements at 211 Hz for both initial polarities. E, f) I-V and P-E measurement with pulse 1 at full amplitude.

By using the approach of inversion domains from the last section and applying it to the results of the unipolar measurements, the 1$^{st}$ and 4$^{th}$ loop can be considered to be in a polarization state with fewer and/or smaller inversion domains. In the first step, the applied field induces the nucleation of further domains at higher electric fields. Therefor growth also takes place at these higher fields even though less energy would be necessary for the domains to expand. This becomes evident from the fact that polarization continues to switch even after reducing the voltage below the initial onset of the switching current. For the 1$^{st}$ cycle this results in a counter clockwise current peak. For the 2$^{nd}$ cycle, where inversion domains are already present from the 1$^{st}$ loop, the domains already grow at lower electric fields, while further nucleation might start at higher electric field. Consequently, $E_c^+$ appears to be reduced by about 10%. This suggests that the majority of the volume can be addressed through domain wall motion from the initial domains. In addition, the overall displacement current peak becomes broader as the domain growth originating from the initial domains is nonetheless likely not able to progress to the whole sample volume. Because similar results were obtained for both polarities of pristine films as well as capacitors that were already switched over 2000 times, most inversion domains appear to be resolved by switching.

### 3.3 Effect of Temperature on Imprint and Inversion Domains

Since the Al$_{1-x}$Sc$_x$N P-E loops still exhibit a noticeable imprint after thousands of cycles and annealing of PZT and HfO was reported to be an effective way to manipulate the center of the hysteresis loop, this section is focused on investigating the effect of temperature on imprint in Al$_{1-x}$Sc$_x$N. [18,29] Therefor, in the first step N- and M-polar capacitors were either annealed for 10 minutes at different temperatures or held for different time intervals at 100 °C and 200 °C. Both temperatures are well below the temperature of phase transition,[4] nonetheless, the different temperatures and time intervals result in varying shifts of imprint. While for fully M-polar capacitors, the negative imprint decreases and even becomes positive for longer time and/or higher temperatures, the imprint in N-polar pristine samples gets more pronounced **(Figure 7 a,b)**. The variation of the hysteresis center along the field axis in the pristine polarization state is 3 to 5 times smaller than the shift of inversely poled capacitors.

In the second step a partially switched sample was heated to 200 °C for 10 min to see whether inversion domains grow, vanish or are otherwise influenced by additional thermal energy. For this, as a reference, a capacitor was first cycled ~50 times with a field amplitude of -4.7 MVcm$^{-1}$. Afterwards a unipolar field of -4.1 MVcm$^{-1}$ was introduced to partially switch the film to M-polarity, followed by a unipolar field of 4.7 MVcm$^{-1}$ to ensure full reversal to N-polarity. For the actual measurement the capacitor was partially switched to M-polarity as described before (-4.1 MVcm$^{-1}$), annealed and then fully switched back to N-polarity with 4.7 MVcm$^{-1}$.
For the partially M-polar sample (Figure 7d), the introduction of temperature resulted in an unexpected shift of the positive switching current towards more positive voltages, even though the great majority of the film can still be expected to be N-polar. Therefore, the shift is opposite to the shift of the fully N-polar films.
The slight increase of I-V area from the preheated to post heated curve at positive fields is a result of the induced shift of the $E_c^-$ by cycling that also increased the switched volume of the negative pre-pulse (see light blue vs. purple curve).

One possible explanation for the imprint shift is a classical defect driven mechanism as proposed for PZT or HfO, where charged defects migrate along the direction of an incompletely screened polarization and therefor change the internal electric field and imprint accordingly.[16,18] With this mechanism, however it is not possible to explain the $E_c$ shift in the partially switched sample (Figure 7d). This mechanism furthermore would not agree with the imprint shift observed after applying unipolar fields as discussed for figure 6. If charged defects were the reason for the change in imprint, the response to the applied field would have to be reversed. A positive field would distribute defects more evenly throughout the sample, reducing the internal imprint. As the opposite is observed and discussed in the previous section, imprint shift by electric fields and annealing would be separate mechanisms.

The more likely explanation, which also agrees to the imprint shift induced by electric fields, is again the stabilization of the present polarization of the films due to structural changes/defects that favor the nucleation or presence of a certain polarity. As the films during deposition grow in the N-polar state at elevated temperature, this polarization state is already stabilized in the process. M-polarity on the other hand is energetically not favored in as deposited films. The introduction of temperature to these films enables the reduction and reorientation of structural defects, making it easier to switch to M-polarity, while at the same time making it harder to switch back to N-polarity. The imprint shift therefor is much stronger for M-polar than for N-polar films. From the partially switched film it becomes evident, that existing minority domains do neither grow nor shrink. However, because minority-polar volume was still stabilized locally during the annealing, it became harder to switch back and the imprint became less negative.

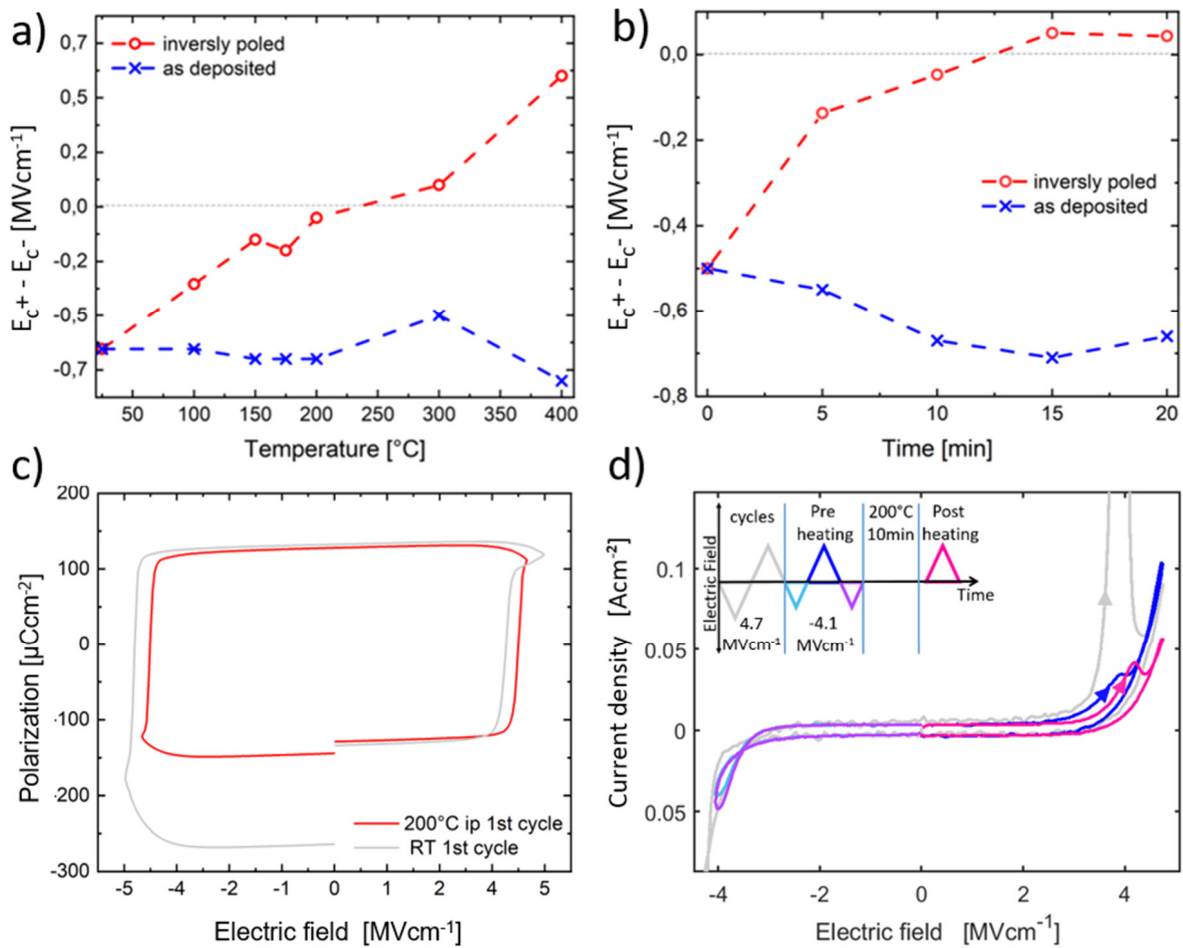

**Figure 7:** *a) Sum of $E_c$ of as deposited and inversely poled $Al_{0.70}Sc_{0.30}N$ capacitors vs. temperature. b) Sum of $E_c$ at 200 °C vs time. c) P-V loops of capacitor at room temperature and heated to 200 °C for 10 min. d) I-V measurement of a capacitor partially switched in negative direction (pre heating) and fully switched afterwards in positive direction (post heating) compared to same measurement without heating. The measurements shown in d) were conducted*

*to confirm the nucleation based explanation. The colors in the pulse train added as an inlet refer to the colors of the curves.*

## 4. Conclusion

Ferroelectric $Al_{1-x}Sc_xN$ films were investigated for wake-up and imprint. While the films experience a wake-up like behavior when cycled with minor loops in the range of the coercive field, they reach full $P_r$ when the voltage amplitude is sufficiently above the coercive field and therefor can be considered wake-up free. We attribute the wake-up like behavior to a shift in the imprint of the films, which is due to the emergence of regions which either are stable inversion domains or at least regions were the as grown minority polarization is favored. For our N-polar films, this results in a pronounced reduction of $E_c^-$ during the initial switching events. This way minor loops turn into major loops and $Al_{1-x}Sc_xN$ films appear to feature wake-up.

Based on these observations, previous studies and SEM imaging of anisotropically etched surfaces, an explanation is proposed for the initial switching cycles. For N-polar films, M-polarity propagates from the top-electrode through the film, likely by domain wall motion. At the bottom electrode N-polar inversion domains remain and serve as initial sites for back-switching while in switched N-polar film regions remain at top electrode which are more favorable for nucleation of M-polar domains, therefor leading to a reduces negative $E_c$ and shift in imprint in consecutive cycles.

Imprint was further investigated under the influence of unipolar fields which were found to destabilize and reduce oppositely oriented inversion domains (or regions, were such inversions are energetically more favorable) present in the film, therefor shifting the hysteresis to higher or lower electric fields.

The general presence of inversion domains was also found to lower the energy required for full polarization inversion in partially switched films such that the coercive field is reduced and the I-V peak broadens in the consecutive cycles. We attribute this to a mechanism were a larger volume of the films becomes addressable to energetically less demanding domain wall motion in the presence of already switched sample volume, in line with our explanation for imprint.

Furthermore temperature was also found to stabilize the polarization present in the heated films and be able to shift the imprint close to 0 MVcm$^{-1}$ for 10 min at 200 °C. This way temperature offers an opportunity to program imprint to a desired value. No domain growth at elevated temperatures was observed.

**Conflict of interest:**

We declare no conflict of interest

**Acknowledgements:**

This Work was funded by the project "ForMikro-SALSA" (Project-ID 16ES1053) from the Federal Ministry of Education and Research
(BMBF) and the Deutsche Forschungsgemeinschaft (DFG, German Research Foundation), Project-ID 458372836.

Received: ((will be filled in by the editorial staff))

**Supporting Information:**

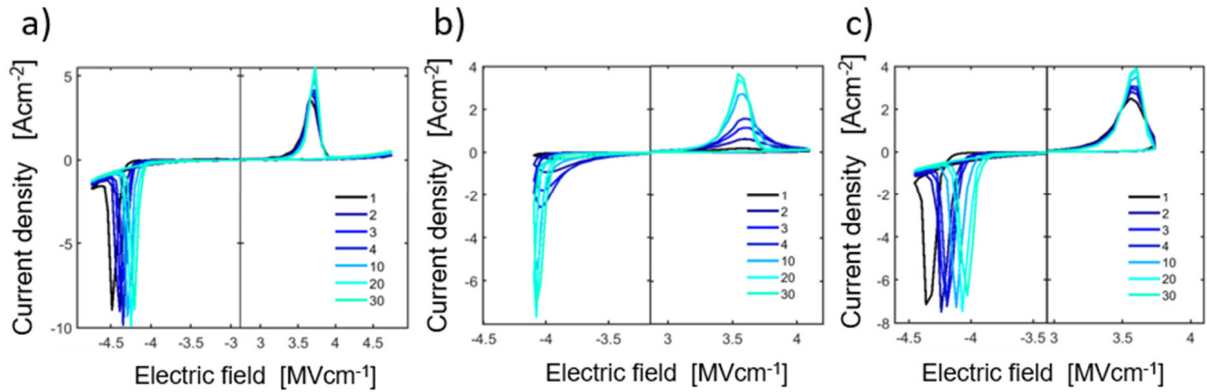

**Figure S1:** *a) I-V loops at 4.7 MVcm$^{-1}$ and a frequency of 211 Hz. b) I-V loops at 4.1 MV/cm and 211Hz without and c) with a bias field of -0.36 MVcm$^{-1}$ for the first 30 cycles on a pristine capacitor on the same sample.*

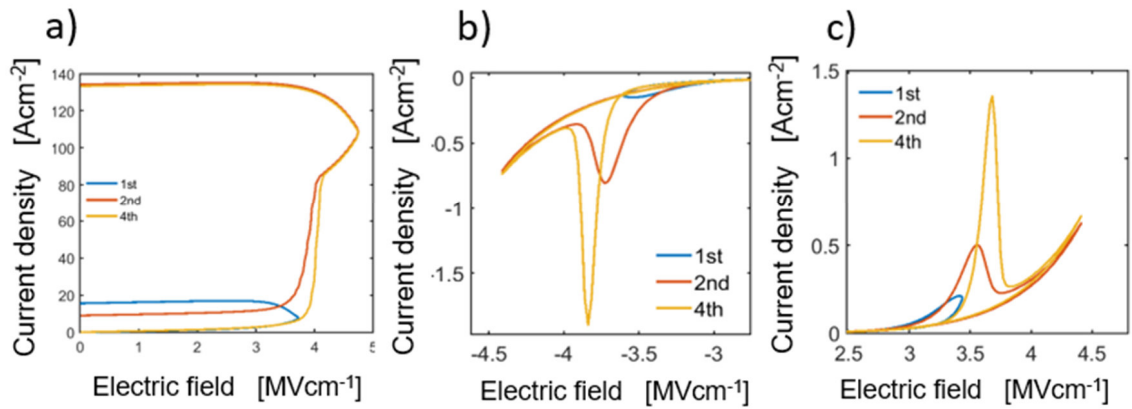

**Figure S2:** *Unipolar measurements. A) P-E measurement corresponding to figure 6c at 211Hz, b,c) I-V measurements at 50Hz.*

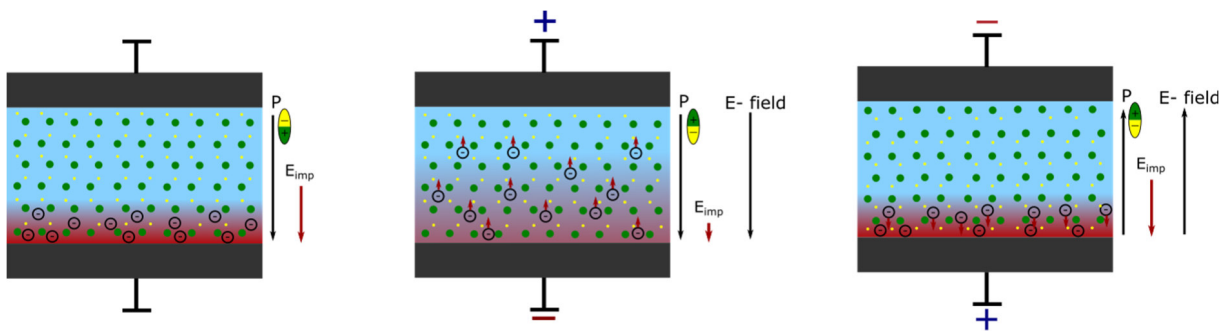

**Figure S3:** *a) Mechanism of charge migration in different external fields with the resulting internal field* E$_{imp}$. *Left: no external field, center: positive external field and right: negative external field*

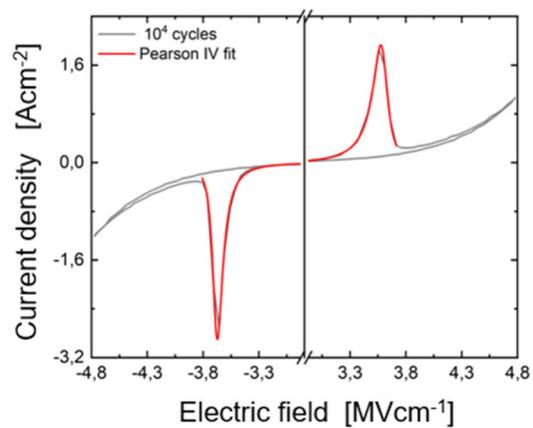

**Figure S4:** *I-V measurement of a capacitor structure with a fit of the switching peaks. The Fit is used to determine the precise coercive field during cycling (see Figure 1)*